\documentclass[%
 reprint,
superscriptaddress,
 amsmath,amssymb,
]{revtex4-2}

\usepackage{graphicx}% Include figure files
\usepackage{dcolumn}% Align table columns on decimal point
\usepackage{bm}% bold math
\usepackage{hyperref}
\hypersetup{hypertex=true,
	colorlinks=true,
	linkcolor=blue,
	anchorcolor=blue,
	urlcolor=blue,
	citecolor=blue}
\begin{document}

\preprint{APS/123-QED}

\title{Tunable Orbital Thermoelectric Transport with Spin-Valley Coupling in Ferromagnetic Transition Metal Dichalcogenides}% Force line breaks with \\
%\thanks{A footnote to the article title}%

\author{Shilei Ji}
\affiliation{School of Science, Jiangsu Provincial Engineering Research Center of Low Dimensional Physics and New Energy, Nanjing University of Posts and Telecommunications, Nanjing 210023, China.}%Lines break automatically or can be forced with \\

\author{Jianping Yang}
\affiliation{School of Science, Jiangsu Provincial Engineering Research Center of Low Dimensional Physics and New Energy, Nanjing University of Posts and Telecommunications, Nanjing 210023, China.}

\author{Li Gao}
\email{iamlgao@njupt.edu.cn}
\affiliation{School of Science, Jiangsu Provincial Engineering Research Center of Low Dimensional Physics and New Energy, Nanjing University of Posts and Telecommunications, Nanjing 210023, China.}
\affiliation{School of Materials Science and Engineering, Nanjing University of Posts and Telecommunications, 9 Wenyuan Road, 210023, Nanjing, China.}

\author{Xing'ao Li}
\email{lxahbmy@126.com}
\affiliation{School of Science, Jiangsu Provincial Engineering Research Center of Low Dimensional Physics and New Energy, Nanjing University of Posts and Telecommunications, Nanjing 210023, China.}
\affiliation{College of science, Zhejiang University of Science and Technology, Hangzhou 310023, China.}

%\collaboration{CLEO Collaboration}%\noaffiliation

\date{\today}% It is always \today, today,
             %  but any date may be explicitly specified

\begin{abstract}
In valleytronic devices, the valley transport of electrons can carry not only charge but also spin angular momentum (SAM) and orbital angular momentum (OAM). However, investigations on thermoelectric transport of OAM manipulated by valley degrees of freedom remain limited. Here, using the ferromagnetic transition metal dichalcogenides RuCl$_2$ as an example, we investigate valley-contrasting Berry curvature and demonstrate its role in generating valley-dependent anomalous and orbital Nernst effects. The thermoelectric transport of OAM is shown to be modulated by intrinsic spin polarization and exhibits characteristics of valley-orbital coupling. Furthermore, we show that spin-valley coupling plays a crucial role in controlling the orbital Nernst effect and distinguishing it from the anomalous Nernst effect. Based on these findings, we propose a thermoelectric transport mechanism for generating pure orbital currents.

\end{abstract}

%\keywords{Suggested keywords}%Use showkeys class option if keyword
                              %display desired
\maketitle

\paragraph*{Introduction.}\!\textemdash{} Valleytronics introduces a novel dimension—the valley degree of freedom—for manipulating Bloch electrons.\cite{RN985, RN1084, RN575, RN635, RN505, RN1070, RN1086} It focuses on the valley degree of freedom of electrons in crystals and its applications in information storage and processing. Two-dimensional materials play a crucial role in the development of valleytronics, particularly following the discovery of transition metal dichalcogenides (TMDCs) materials such as MoS$_2$.\cite{RN738, RN557, RN305, RN14, RN575, RN1028} These materials exhibit two energy extrema in momentum space, corresponding to the valley degrees of freedom, which can be effectively manipulated using electrical and optical techniques.\cite{RN1027, RN979, RN830, RN1021, RN1025}

The valley degree of freedom is a manifestation of the Bloch electronic structure and can be detected through carriers transport.\cite{RN985, RN575, RN635, RN1028, RN636} The wavefunction of Bloch electrons can be expressed as $ \psi_{n\mathbf{k}} = e^{i\mathbf{k} \cdot \mathbf{r}} u_{n\mathbf{k}} $, where $ n $ denotes the band index, encoding information about the spin and atomic orbital characteristics.\cite{RN1089, RN778} Therefore, the motion of valley-scattered electrons is influenced by the spin and orbital degrees of freedom.\cite{RN733, RN1084, RN944, RN947, RN575} The transport of Bloch electrons is related to the valley-contrasting Berry curvature, which can be regarded as an effective magnetic field in momentum space.\cite{RN734, RN733, RN736, RN946, RN735, RN730} Under a temperature gradient, the Berry curvature together with valley degree of freedom leads to the anomalous Nernst effect (ANE), influencing the transverse motion of carriers and generating a transverse current.\cite{RN1016, RN1029, RN1028,RN1018, RN960, RN1025, RN694} Moreover, in TMDCs, the ANE induces pure spin and valley currents, which can be regarded as spin and valley Nernst effects, respectively.\cite{RN1018, RN1028} The valley degree of freedom in Bloch states is associated with the orbital magnetic moment, meaning that thermoelectric transport driven by Berry curvature is intrinsically orbital-polarized.\cite{RN1020, RN960, RN1025} However, research on thermoelectric transport involving OAM is still scarce, especially concerning the valley-dependent orbital Nernst effect (ONE), which remains largely unexplored.\cite{RN1017, RN739, RN1009}

In this work, we take the TMDC material RuCl$_{2}$ as an example to fill this gap by proposing a spin-valley coupling mechanism that manipulates the ONE and leads to the switching between pure orbital current and orbital-polarized current. The breaking of inversion symmetry gives rise to Berry curvature, while the breaking of both time-reversal and inversion symmetries induces intrinsic spin-valley coupling. Through $k \cdot p$ model analysis and first-principles calculations, we show that TMDCs like RuCl$_{2}$ exhibit intrinsic ANE and ONE, which arise from valley-contrasting Berry curvature. In the following, we investigate the manipulation of valley-resolved ONE by spin polarization, which induces pure orbital currents. By examining the influence of spin-valley coupling on the Berry curvature, we establish its connection with both the ONE and ANE.

\begin{figure}[t]
	\centering
	\includegraphics[width=\linewidth]{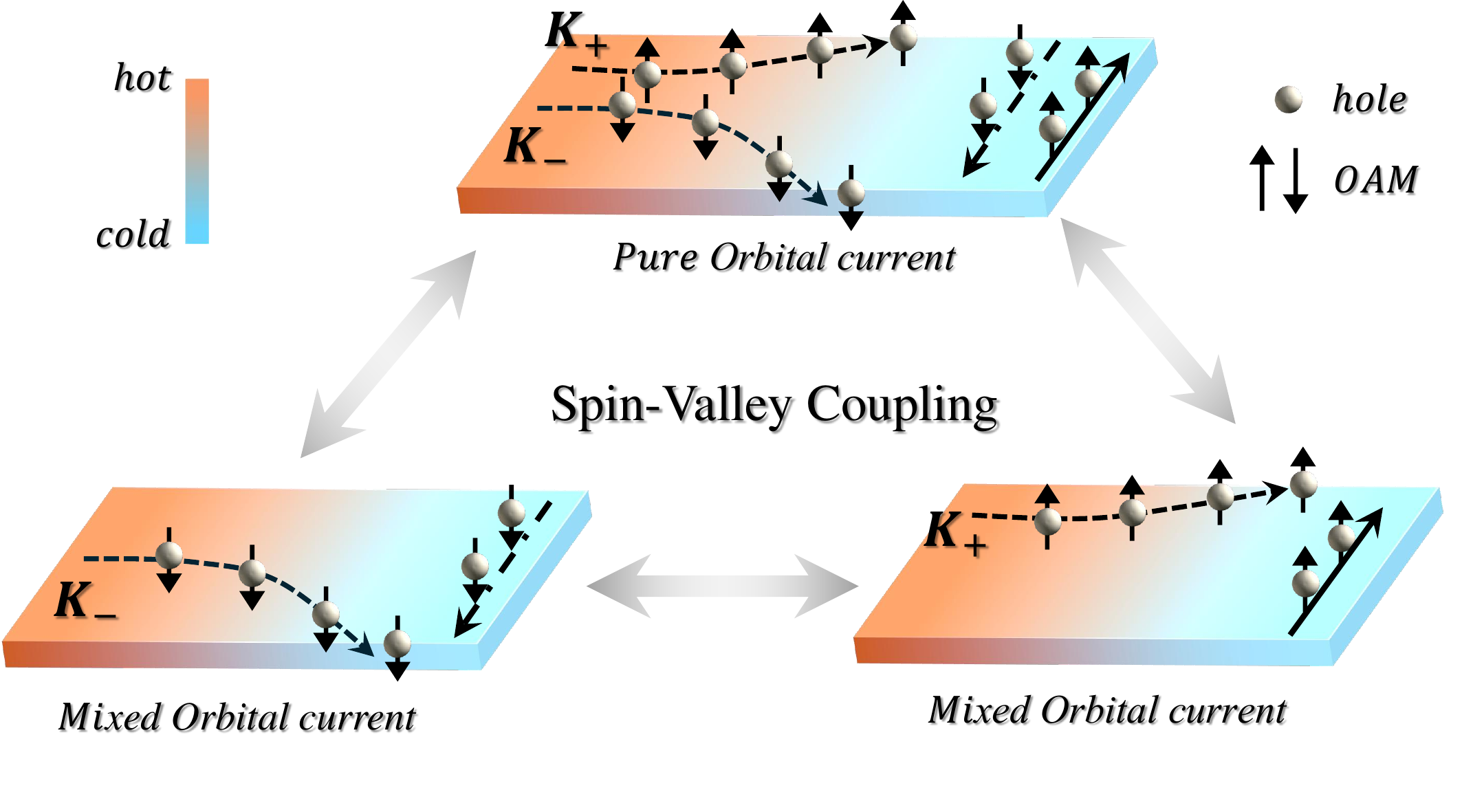}
	\caption{ Schematic of the spin-valley coupling tunbale orbital thermoelectric transport. When the temperature gradient is applied transversely to the sample, the anomalous and orbital Nernst effects occur on the $K_+$ and $K_-$ valley. The gray balls represent holes and up/down arrows are the orbital degree of freedom with $\langle \hat{L}_z\rangle=+2 / -2 \hbar$. Moreover, the pure orbital current indicates the accumulation of orbital angular momentum (OAM) at both edges of the sample without electronic potential difference. The mixed orbital current indicates that both edges of the sample exits OAM and electronic potential difference.
	}
	\label{fgr1}
\end{figure}

\paragraph*{Method.}\!\textemdash{} The orbital Berry curvature can be obtained by kubo formula\cite{RN746, RN745, RN466},
\begin{equation}\label{eqn-5}
	\Omega_{n}^{\hat{L}_z}(\textbf{\textit{k}}) = 
	-2\hbar \sum_{n\neq n'}
	\frac{{\rm Im}\langle \psi_{n\textbf{\textit{k}}}|\hat{v}_x|\psi_{n'\textbf{\textit{k}}}\rangle 
		\langle \psi_{n'\textbf{\textit{k}}}|\hat{J}_y| \psi_{n\textbf{\textit{k}}}\rangle}
	{(E_{n'} - E_{n})^2} 
\end{equation} 
Where $\hat{v}_i$ $(i = x, y)$ is the velocity operator along the $k_i$ direction and the OAM operator $\hat{J}_y$ is defined as $\hat{J}_y = \frac{1}{2}(\hat{v}_y\hat{L}_z+\hat{L}_z\hat{v}_y)$. $\hat{L}_z$ is the $z$ components of the OAM operator. The ONE can be written as

\begin{equation}
	\alpha_{xy}^{\hat{L}_z} = 
	\frac{e k_B}{\hbar} \sum_{n}
	\int_{BZ}
	\frac{d^2k}{(2\pi)^2} \Omega_{n}^{\hat{L}_z}(\textbf{\textit{k}})
	S_{n}(\textbf{\textit{k}}),
\end{equation} 
where $S_n(\textbf{\textit{k}})=-f_{n\textbf{\textit{k}}} \rm{ln}  f_{n\textbf{\textit{k}}}-(1-f_{n\textbf{\textit{k}}}) \rm{ln} (1-f_{n\textbf{\textit{k}}})$, and $f_{n\textbf{\textit{k}}}$ is the Fermi-Dirac distribution. $k_B$ is the Boltzmann constant and $e$ is the magnitude of the electron charge. The Berry curvature $	\Omega_{n}^{z}(\textbf{\textit{k}})$ and ANE $	\alpha_{xy}^{z}$ can be obtained by replacing $\hat{J}_y$ to $\hat{v}_y$. 

\paragraph*{Model arguments.}\!\textemdash{}Starting with a two-band $k \cdot p$ model that simultaneously breaks both time-reversal and inversion symmetries, with the basis functions are  $\psi_c^{\tau} = |d_{z^2} \rangle $ and $\psi_v^{\tau} =(|d_{x^2-y^2} \rangle + i\tau|d_{xy} \rangle) /\sqrt{2}  $ for the conduction (\textit{c}) and valance (\textit{v}) bands, where $\tau=+1(-1)$ represents the $K_+$($K_-$) valley.\cite{RN733, RN944, RN792, RN729, RN960} This $k \cdot p$ model was previously proposed to explain the origin of the OHE and topological phase transition in two-dimensional TMDCs, which has the spin-valley coupling at two valleys. The effective Hamiltonian $H$ can be interpreted as a Dirac cone model with broken time-reversal and inversion symmetries:

\begin{equation}\label{eqn-3}
	\begin{aligned}
		H = 
		v_F (\tau \hat{\sigma}_xk_x+\hat{\sigma}_yk_y) +\dfrac{\Delta}{2} \hat{\sigma}_{z} - \tau \cos\theta_z \lambda_v \frac{\hat{\sigma}_{z}-1}{2}.
	\end{aligned}
\end{equation}

Here, $v_F$ is the massless Fermi velocity of the Dirac electrons, $\hat{\sigma}_i (i = 0, x, y, z)$ is Pauli matrice for pseudospin, and $\theta_z$ is the rotation angle of the spin-polarization along \textit{z}-axis. In addition,  the band gap of the \textit{c} and \textit{v} bands is represented by $\Delta$, and $\lambda_v$ is the SOC parameter for the \textit{v} band. 

By diagonalizing the Hamiltonian $H$, the eigenvalues are expressed as: $ E_n =\frac{1}{2}(\tau \cos\theta_z \lambda_v - n\sqrt{4v_F^2k^2+\Delta_{\tau}^2}) $, where $\Delta_{\tau}= \Delta-\tau \cos\theta_z \lambda_v$, and $k^2$ is equal to $k_x^2+k_y^2$. The subscript $n=-1(+1)$ is the band index for $c(v)$ band.
The Berry curvature $\Omega_{n,\tau}^{z}(\textbf{\textit{k}})$ and orbital Berry curvature $\Omega_{n,\tau}^{\hat{L}_z}(\textbf{\textit{k}})$ at the valleys can be written as
\begin{equation}
	\begin{aligned}
		\Omega_{n,\tau}^{z}(\textbf{\textit{k}}) = \tau \times \Omega_{n,\tau}^{\hat{L}_z}(\textbf{\textit{k}})
		= \tau n \dfrac{ v_F^2 \Delta_{\tau} }{(4v_F^2k^2
			+\Delta_{\tau}^2)^{3/2}}
	\end{aligned}
\end{equation} 
 At low temperatures $T$, the Eq.(2) can be reduced to the Mott relation\cite{RN1025},
\begin{equation}
	\alpha_{n,\tau}^{z}= -e \tau \times \alpha_{n,\tau}^{\hat{L}_z}=\tau n \dfrac{e\pi^2 k_B^2 T}{12 h} \dfrac{\Delta_{\tau}}{E_F^2},
\end{equation}
where $	\alpha_{n,\tau}^{z}$ and $\alpha_{n,\tau}^{\hat{L}_z}$ are the anomalous and orbital Nernst conductivity, respectively. $E_F$ is the Fermi level. Here we discuss the case of band index $n=+1$, which is opposite to that of $n=-1$.  The equation only holds at the time of $|E_F|>|\Delta_{\tau}|$, which means that the Fermi level crosses the $v$ bands. The sign of $\alpha_{n,\tau}^{z}$ is related to the valley index, indicating valley-contrasting charge current. In contrast, the sign of $\alpha_{n,\tau}^{\hat{L}_z}$ is identical for both valleys, signifying the emergence of orbital-polarized current. To characterize spin-valley-dependent ONE, we define a orbital-polarized current mixing factor as 
\begin{equation}
\eta=\frac{\alpha_{K_+}^{\hat{L}_z}-\alpha_{K_-}^{\hat{L}_z}}{\alpha_{K_+}^{\hat{L}_z}+\alpha_{K_-}^{\hat{L}_z}}\times 100\%,
\end{equation}
where $\eta=+1(-1)$ represent the total ONE is contributed by $K_+(K_-)$ valley. Since the OAM at  $K_{\pm}$ valley is $\pm 2\hbar$, the opposite-sign of Orbital current can be obtained by $\eta=\pm1$. 

Here we assume $E_F$ lies between two valleys, $\Delta-|\lambda_v\cos\theta_z|\leq|E_F|\leq\Delta+|\lambda_v\cos\theta_z|$, demonstrating that the non-zero ONC exists only in the valley near the Fermi level. As the spin polarization direction $\theta_z$ changes, the value of $\eta$ switches between $-100\%$, 0\%, and $+100\%$. When $\theta_z$ is oriented along the [001] direction, the Fermi level crosses only the $K_+$ valley, resulting in $\eta=+100\%$. This indicates that the total ONE, only contributed by $K_+$ valley, generates an orbital current with angular momentum $\langle \hat{L}_z \rangle=+2\hbar$. At this time, the ANE appears at the $K_+$ valley, leading to the mixed orbital current.  When $\theta_z$ is oriented along the $[00\bar{1}]$ direction, the result is reversed. In addition, $\theta_z=\pi/2$, lying in the $xy$ plane, causes the $\alpha_{K_+}^{\hat{L}_z}=\alpha_{K_-}^{\hat{L}_z}$, resulting in ONE appearing simultaneously in two valleys and generating the pure orbital current. Going back to conditional $E_F$, we find that the value of the $E_F$ depends on $\Delta\pm\lambda_v\cos\theta_z$, and the second term is determined by spin polarization and spin-orbit coupling. In valleytronics, $\lambda_v\cos\theta_z$ is referred to as spin-valley coupling.\cite{RN641, RN570, RN630, RN535, RN721} Therefore, the orbital polarized current is mainly manipulated by spin-valley coupling.

\paragraph*{Realistic systems.}\!\textemdash{}In order to verify the spin-valley coupling tunable ONE in TMDCs, we choose 1L RuCl$_2$ as an example by First-principle calculations. RuCl$_2$ has a crystal structure of hexagonal honeycomb with $D_{3h}$ space group, which can split $4d$ orbitals into $A_1$($d_{z^2}$), $E_1(d_{xy},d_{x^2-y^2})$ and $E_2(d_{xz},d_{yz})$ orbitals.\cite{RN946, RN1090, RN593, RN713} As is shown in Fig. 2(b), both two valleys near the Fermi level are spin-polarized with the SAM $\langle\hat{S}_z \rangle= -1/2 \hbar$, demonstrating the spin-polarization angle is $\theta_z=\pi$. Combining the previous $k\cdot p$ model, we can obtain the spin-valley coupling effect between valley polarization $\Delta_v$ and spin polarization $\theta_z$ as $\lambda_{SOC}=2\lambda_v \cos \theta_z$. Since the $E_1$ orbital is localized within the $v$ band of valley, the magnetic quantum number ($m_l$) contributes a nonzero valley OAM, as shown in Fig. 2(c). The projection of OAM can be obtained using the expectation value equation, $\langle \hat{L}_{z,\textbf{\textit{k}}} \rangle= \langle \psi_{\textbf{\textit{k}}}|  \hat{L}_z |\psi_{\textbf{\textit{k}}} \rangle$.\cite{RN804, RN1006, RN953} Despite the breaking of time-reversal and inversion symmetries, a valley-contrasting OAM ($\langle \hat{L}_{z}^\tau \rangle = 2\tau \hbar$) remains present at the $v$ band. The valley-contrasting OAM in the $v$ band demonstrates intrinsic valley-orbital coupling in RuCl$_2$, $\langle \hat{L}_z \rangle = \tau m_l \hbar $. By tuning the valley degrees of freedom, one can manipulate the orbital degrees of freedom. Furthermore, spin-valley coupling in RuCl$_2$ allows for valley manipulation via spin polarization, enabling effective switching of the OAM.

%In addition, the giant valley polarization at the $v$ band originates from the SOC effect between out-of-plane spin polarization and the $E_1$ orbital with magnetic quantum number $m_l=\pm2$, which produces a spin-valley coupling effect. 

%The AHE and OHE are effective methods to detect the transport of electron charge and OAM. 
The breaking of inversion symmetry leads to the presence of opposite-sign Berry curvature in the Dirac valleys, which can be interpreted as an effective magnetic field in momentum space, inducing an anomalous velocity of charge carriers.\cite{RN508, RN853, RN575, RN635, RN597} The sign of Berry curvature, along with the type of charge carriers, dictates the direction of anomalous velocity. In a crystal, the motion of charge carriers not only conveys SAM but also carries OAM derived from the electronic structure, resulting in OAM accumulation at the sample edges.\cite{RN804, RN1006, RN944} To distinguish between the contributions of OAM and charge carriers, we incorporate the OAM matrix into the Berry curvature equation, as outlined in Eq. 1. The calculated results for the charge Berry curvature and orbital Berry curvature of RuCl$_2$ are illustrated in Fig. 2(e). Since charge carrier transport primarily arises from transitions between valleys, these two valleys contribute most significantly to both the Berry curvature and the orbital Berry curvature. Under an in-plane electric field, the inter-band transition holes exhibit two key features: (i) opposite anomalous velocities due to Berry curvature induced by inversion symmetry breaking, and (ii) valley-contrasting OAM originating from the orbital electronic structure. Consequently, the orbital Berry curvature has the same sign in both valleys, resulting in the orbital polarized current perpendicular to the electric field. In contrast, the charge current is suppressed because the Berry curvature in each valley has opposite signs. The corresponding Hall conductivities $\sigma_{xy} = 0 e^2/h$ and $\sigma_{xy}^{\hat{L}_z} = 1.25 e/2\pi$ within the band gap, shown in Fig. S1 \cite{RNSI}, confirm the existence of pure orbital current. While the pure orbital current induced by in-plane Electronic field is hard to manipulate. Since the OAM channel of the opposite sign exists within the band gap of valley, the valley state is difficult to be suppressed by the external field.

\begin{figure}[t]
	\centering
	\includegraphics[width=\linewidth]{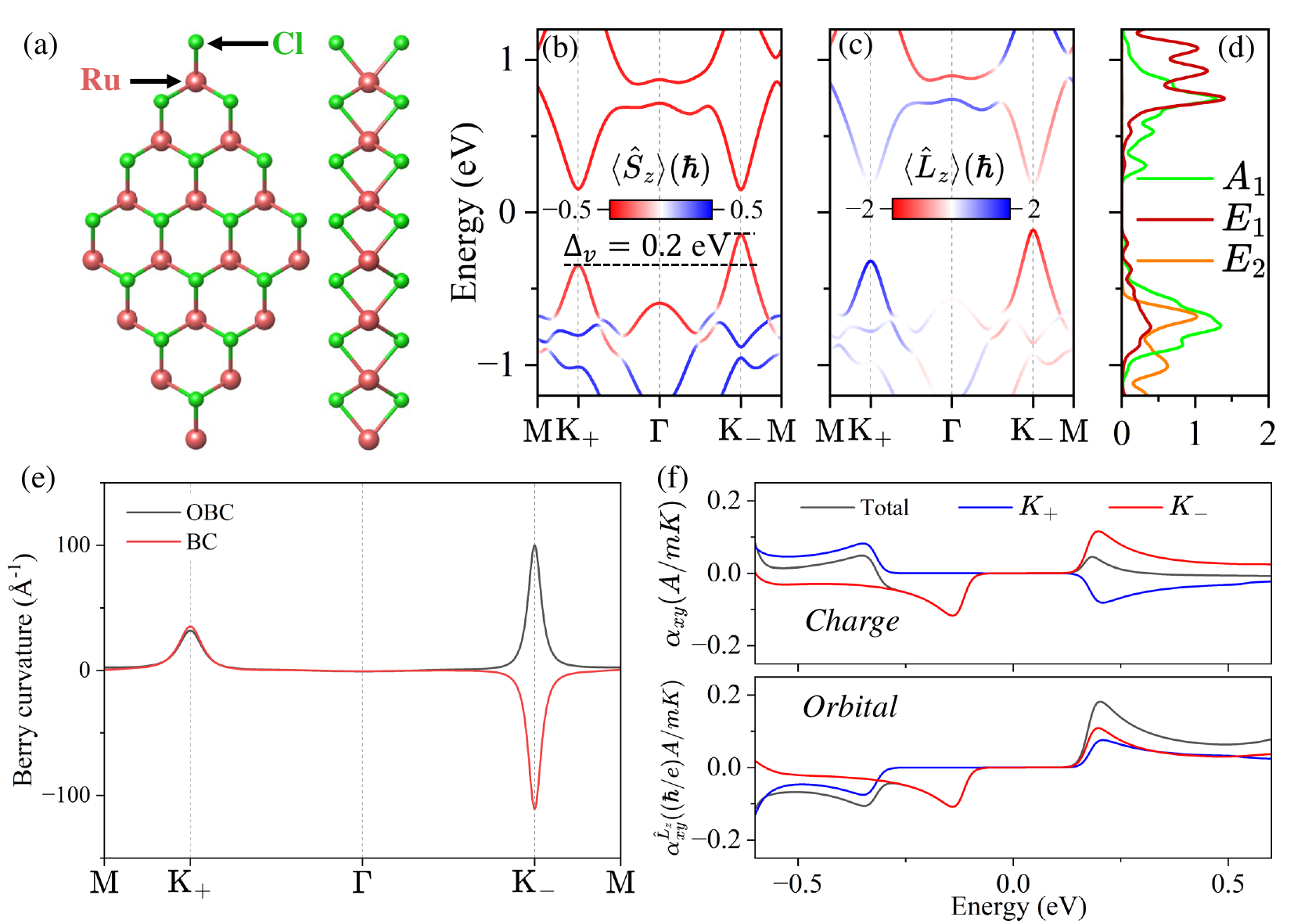}
	\caption{ (a) The atomic structure of 1L RuCl$_2$. Left and right panels are the top and side view, respectively. (b) The spin and (c) orbital band structure for 1L RuCl$_2$, where $\langle \hat{S}_{z}\rangle $ and $\langle \hat{L}_{z}\rangle $ represent the spin and orbital angular moment. In addition, the valley splitting $]Delta_v$ is equal to 0.2 eV. (d) The density of states of RuCl$_2$, where  $A_1$, $E_1$ and $E_2$ orbitals represent $d_{z^2}$, $d_{xy}+d_{x^2+y^2}$ and $d_{xz}+d_{yz}$ orbitals. (e) The Berry curvature and orbital Berry curvature are plotted in red and black lines. (f) Up and down panels are the anomalous and orbital Nernst conductivity. The black, blue and red lines are the corresponding Nernst conductivity for the total Brillouin zone, $K_+$ valley and $K_-$ valley, respectively.
	}
	\label{fgr2}
\end{figure}

\paragraph*{Thermoelectric effect.}\!\textemdash{}Meanwhile, the Seebeck effect, driven by applied temperature gradient, can substitute for an external electric field, enabling the thermoelectric effect.\cite{RN1016} Due to the influence of Berry curvature, an electric potential difference develops across the sample perpendicular to the temperature gradient, known as ANE.\cite{RN1018, RN1020} In Fig. 2(f), the magnitude of ANE for RuCl$_2$ are characterized by ANC. By filtering contributions within the first Brillouin zone, the valley-dependent ANE can be determined. Here, we focus on the $v$ band. The opposite signs of ANC peaks for the two valleys consistent with our $k \cdot p$ model analysis, indicating that the temperature gradient induces the valley-contrasting anomalous velocity for the holes. 

Subsequently, we investigate the accumulation of OAM under the temperature gradient and calculate the ONE in RuCl$_2$, with the ONC displayed in the lower panel of Fig. 2(f). The ONE exhibits peaks of the same sign around the two valleys, which is in contrast to the results observed for ANE. This observation suggests that the holes excited in the two valleys have opposite OAM, leading to the generation of orbital-polarized current. Under the influence of gate voltage, the ONE can effectively generate the OAM of $-2\hbar$ at the $ K_-$ valley while neglecting the OAM contribution from the $K_+$ valley. This behavior differs from the OHE, where both valleys consistently produce opposite OAM contributions. In addition, we calculate the orbital-polarized current mixing factor $\eta$ as a function of energy in Fig. S2 \cite{RNSI}.  Near the $K_-$ valley, the ONE produces a 100\% polarized orbital current with $\langle \hat{L}_z \rangle = -2 \hbar$. At lower energy levels, however, the value of $\eta$ fluctuates between $-100\%$ and $+100\%$, indicating that the orbital current generated by the ONE at these energies involves contributions from both valleys. It is evident that the $\eta=-100\%$ for the $ K_-$ valley remains robust with temperature. While the ONE in both valleys strengthens as temperature increases, the valley polarization induced by SOC effect separates the $K_+$ and $K_-$ valleys. As a result, near $-0.1$ eV, the ONE contribution arises exclusively from a single valley, leading to a 100\% orbital polarized current.

\begin{figure}[tb]
	\centering
	\includegraphics[width=\linewidth]{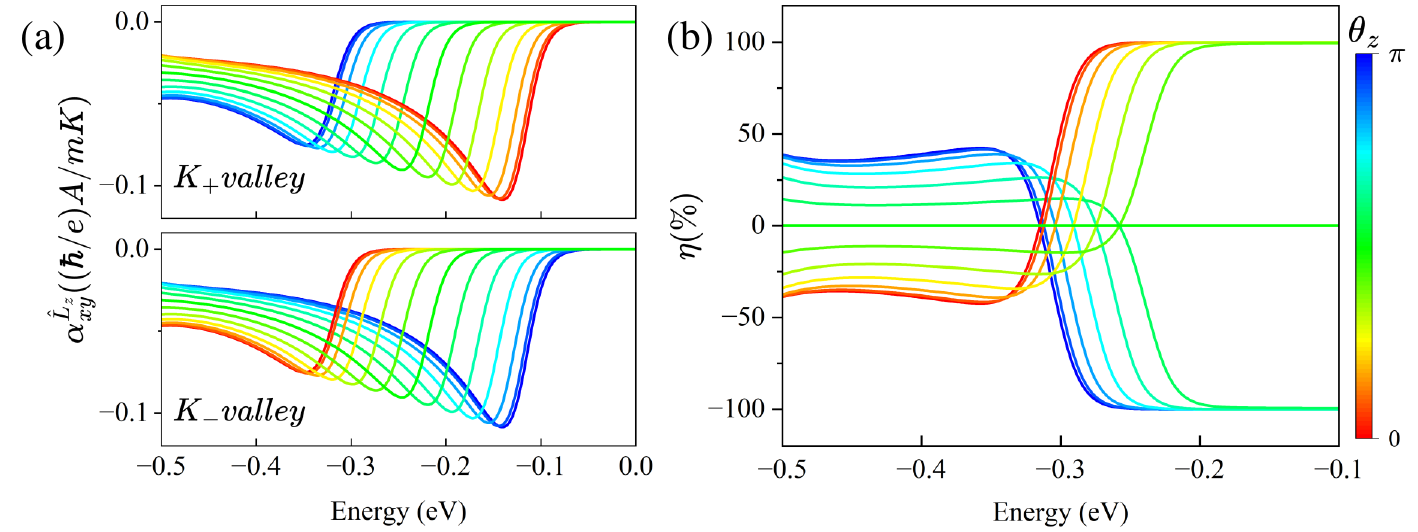}
	\caption{ (a) The valley-resolved orbital Nernst conductivity of $K_+$ and $K_-$ valleys, (b) orbital-polarized current mixing factor for spin-polarization angle.
	}
	\label{fgr3}
\end{figure}
To explore the influence of spin polarization on the ONE, we incorporate spin polarization into our calculations in Fig. 3. In this work, the spin polarization angle $\theta_z$ for RuCl$_2$ is set to $\theta_z=\pi$. As $\theta_z$ decreases from $\pi$ to 0, the Berry curvature maintains its peaks in both valleys, indicating that carriers excited in the valleys retain an anomalous velocity, thereby inducing an intrinsic ONE. As shown in Fig. 3(a), spin polarization manipulates the valley polarization, causing the ONE peaks in the two valleys to shift in opposite directions, which is correspond to the conclusion discussed by $k \cdot p$ model. As the spin-polarization angle changes, the orbital-polarized current mixing factor also varies. When $\theta_z = \pi$, the ONE near the Fermi level is primarily contributed by the $K_-$ valley, resulting in a $100\%$ orbital-polarized current with $-2\hbar$ OAM. As the angle decreases, valley polarization is reduced, and the ONE contributions from the two valleys gradually converge. As shown in Fig. 3(b), this leads to sign-reversal in $\eta$; in particular, at $\theta_z = \pi/2$, the ONE contributions from both valleys are equal, resulting in the accumulation of equal magnitude of holes with opposite-sign OAM on opposite side of RuCl$_2$. This configuration produces a purely orbital current without electrical difference.

\begin{figure}[t!]
	\centering
	\includegraphics[width=\linewidth]{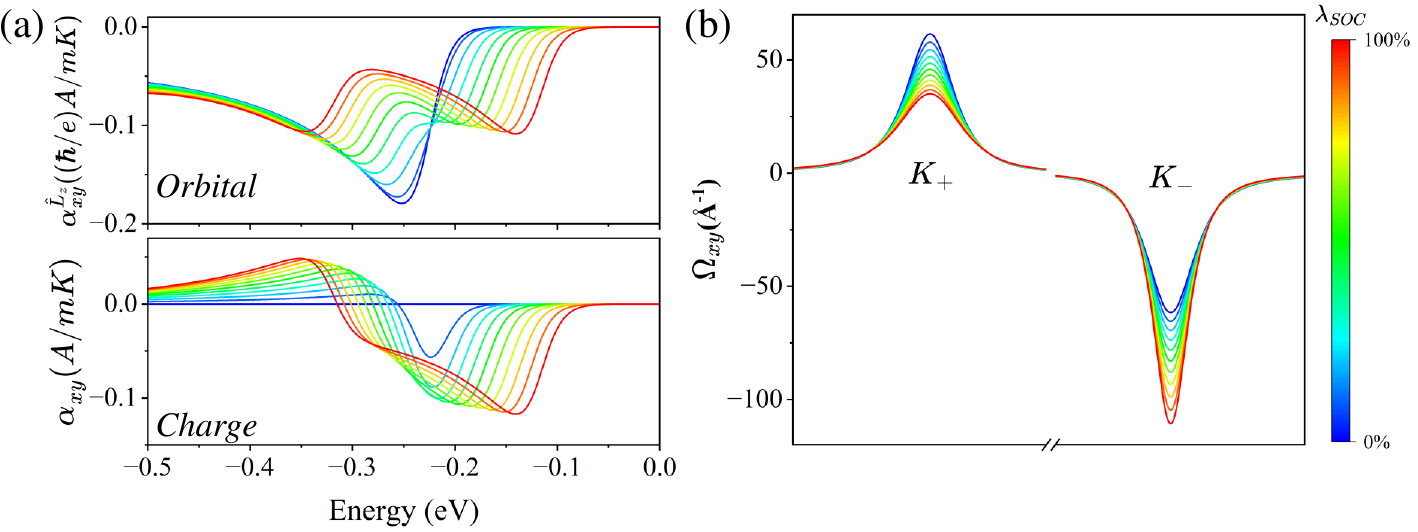}
	\caption{ (a) The orbital, anomalous Nernst conductivity, (b) Berry curvature for $\lambda_{SOC}$, where $\lambda_{SOC}$ is the strength of spin-valley coupling, $\lambda_{SOC}= \cos \theta_z \lambda_v$.
	}
	\label{fgr4}
\end{figure}

\paragraph*{Discussion.}\!\textemdash{}The direction of spin-polarization can be effectively controlled by the magnetic field, electronic field, and magnetic proximity effect.\cite{RN44, RN230, RN418} By applying an external field to manipulate the spin-polarization angle, the valley degrees of freedom can be effectively switched through the SOC effect.\cite{RN557, RN14} Additionally, valley-orbital coupling allows for the occupation of opposite OAM in the two valleys. Under the application of an effective temperature gradient, the breaking of time-reversal symmetry and inversion symmetry in TMDs leads to the emergence of a Berry curvature field, resulting in hole accumulation at both ends of the system. With the rotation of spin polarization, there are three types of orbital-polarized current. (i) When $\eta = + 100\%$, TMDCs generate the charge current with $+2\hbar$ OAM, which can be regarded as mixed orbital current. (ii) When $\eta = 0\%$, TMDs generate the pure orbital current without charge current.  (iii) When $\eta = - 100\%$, TMDs generate the mixed orbital current, which is opposite to case (i). 

The SOC effect together with spin-polarization leads to the intrinsic valley-polarization, as revealed in our $k \cdot p$ model analysis, $\Delta_{\tau}=\Delta - \tau \cos\theta_z \lambda_v$. When the second term $\lambda_{SOC}= \cos\theta_z \lambda_v$ equals zero, the two valleys become degenerate, resulting in an ANE of $\alpha_{K_+}^z= -\alpha_{K_-}^z$ for the two valleys. As a result, the total ANE for the system vanishes. However, under this condition, the ONE for both valleys remains equal, yielding  $\alpha_{K_+}^{\hat{L}_z}= \alpha_{K_-}^{\hat{L}_z}$, confirming that the intrinsic ONE for the system can arise without the SOC and spin-polarization. 
%
%During the thermoelectronic transport, such as Nernst effect, the Berry curvature introduces an anomalous velocity component for charge carriers, leading to charge deflection. When the spin polarization lies in the plane, equal charge accumulation occurs at both ends of the system, causing the ANE to vanish. However, in this configuration, the ONE  reaches a maximum. The SOC effect together with spin-polarization induces an imbalance in carrier deflection between the two valleys, leading to intrinsic AHE and ANE. As revealed in our $k \cdot p$ model analysis, both effects are directly related to the parameter $\Delta_{\tau}=\Delta - \tau \cos\theta_z \lambda_v$. When the second term equals zero, the two valleys become degenerate, resulting in an ANE of $\alpha_{K_+}^z= -\alpha_{K_-}^z$ for the two valleys. As a result, the total ANE for the system vanishes. However, under this condition, the ONE for both valleys remains equal, yielding  $\alpha_{K_+}^{\hat{L}_z}= \alpha_{K_-}^{\hat{L}_z}$, confirming that the intrinsic ONE for the system can arise without the SOC and spin-polarization. 

When $\lambda_{SOC}$ is weak, the dependencies of ANE and ONE on $\lambda_{SOC}$ show opposite trends: as $\lambda_{SOC}$ strength increases, ANE rises while ONE rapidly decreases. However, as $\lambda_{SOC}$ becomes stronger, both ANE and ONE increase together. As shown in Fig. 4, the Berry curvature peaks in the two valleys are robust, with the sign of each remaining unchanged regardless of $\lambda_{SOC}$. The opposite signs of the Berry curvature in the two valleys ensure that carriers exhibit opposite anomalous velocities. Notably, SOC and spin polarization jointly influence inter-valley carrier dynamics. When both effects are accounted for, an imbalance arises in the deflection of carriers between valleys, giving rise to an enhancement of ANE with the $\lambda_{SOC}$ increasing. At this point, the decrease in ONE results from valley splitting driven by the combined effects, leading to an energy-scale separation of the ONE contributions from each valley. The total ONE of the system primarily results from the combined contributions of the ONE from both valleys. As the band gap difference between the two valleys gradually increases, SOC and spin polarization disrupt the carrier balance at both ends of the system, affecting the accumulation of OAM and generating an orbital-polarized current, which reduces the total ONE peak. When the total ONE peak splits into two valley-resolved peaks, the total ONE contribution shifts from a combined effect of both valleys to that of a single valley. Due to the inter-band transition, SOC and spin polarization enhance the number of excited carriers in the valleys near the Fermi level, leading to an increase in the total ONE.

In experiments, the SOC strength of the crystal is related to the type of atom, which is hard to manipulate. The magnitude of the spin-valley coupling is determined by the direction of spin polarization. In addition, the orbital polarized current can be experimental detected by (i) the inverse orbital Hall effect\cite{RN956}, analogous to the inverse spin Hall effect, involves the conversion of orbital thermoelectric currents into electrical signals. (ii) Orbital torque\cite{RN806}: Orbital-polarized current is injected into an adjacent ferromagnetic layer, where the SOC effect generates a spin torque that acts on the ferromagnet. The sign of OAM is detected through the spin polarization direction of the ferromagnetic material. (iii) Magneto-optical Kerr effect\cite{RN776}: The orbital-polarized current can be obtained by detecting the Kerr rotation angle of the reflected light at the crystal edge.

\paragraph*{Conclusion.}\!\textemdash{} In summary, we establish that orbital thermoelectric transport is fundamentally driven by the valley-contrasting Berry curvature, which plays a key role in carriers transport. Furthermore, we show that this mechanism is intricately modulated by spin-valley coupling, which introduces an additional degree of control over the thermoelectric response. The ONE originates as a direct consequence of the ANE, both driven by the Berry curvature. Holes acquire OAM that is intrinsically tied to valley index, giving rise to an orbital-polarized current. When spin-valley coupling is absent, the Berry curvature fields in the two valleys are equal in magnitude but opposite in sign. As a result, the holes accumulate symmetrically at opposite edges of the crystal, carrying OAMs of opposite signs. This symmetry leads to the generation of a pure orbital current. The spin polarization angle affects the wavefunction of holes in the valley, thereby modulating the strength of spin-valley coupling and Berry curvature. Our findings demonstrate that spin polarization can effectively control the accumulation of opposite-sign OAM at the edges of the crystal, resulting in an orbital-polarized current. The spin-valley coupling, as a tunable parameter, governs the ONE, providing a theoretical foundation for the thermoelectric transport of OAM. This establishes spin-valley coupling as a crucial mechanism for manipulating and detecting orbital degrees of freedom in valleytronic systems.

\begin{acknowledgments}
	We acknowledge the fundings from National Natural Science Foundation of China (Grant No. 51872145), and Jiangsu Province Key R\&D Program (No. BK20232044).
\end{acknowledgments}

\bibliography{apssamp}% Produces the bibliography via BibTeX.

\end{document}